\documentclass[english,aps,10pt,prl, floatfix, notitlepage, superscriptaddress,a4paper,
reprint]{revtex4-2}

\newif\iftwocol
\twocoltrue
\usepackage[normalem]{ulem}

\usepackage{lmodern}
\usepackage{amsmath}
\usepackage{amsfonts}
\usepackage{dsfont}
\usepackage{graphicx}
\usepackage{xcolor}
\usepackage{braket}
\usepackage[format=plain,justification=centerlast]{caption}

\usepackage{txfonts}
\let \ns \! 
\renewcommand{\!}{}

\newcommand*{\tcbr}{\nonumber \\ &}

\usepackage{mathtools}
\usepackage{bm}

\usepackage{booktabs}

\usepackage{wasysym}

\usepackage[unicode=true,pdfusetitle,
bookmarks=true,bookmarksnumbered=true,bookmarksopen=true,bookmarksopenlevel=2,
breaklinks=false,pdfborder={0 0 0},backref=false,colorlinks=true]{hyperref}
\usepackage{bookmark}

\renewcommand{\d}{\dagger}
\providecommand*{\hc}{\mathrm{H.c.}}
\renewcommand*{\vec}[1]{{\bm{\mathrm{#1}}}}

\providecommand*{\xop}[1]{{\hat{\chi}_{#1}^{\vphantom{\d}}}}
\providecommand*{\xdop}[1]{\hat{\chi}_{#1}^{\d}}

\providecommand*{\sop}[1]{{\hat{S}_{#1}}}

\providecommand*{\aop}[1]{{\hat{a}_{#1}^{\vphantom{\d}}}}
\providecommand*{\adop}[1]{\hat{a}_{#1}^{\d}}

\providecommand*{\bop}[1]{{\hat{b}_{#1}^{\vphantom{\d}}}}
\providecommand*{\bdop}[1]{\hat{b}_{#1}^{\d}}

\providecommand*{\hop}[1]{{\hat{h}_{#1}^{\vphantom{\d}}}}
\providecommand*{\hdop}[1]{\hat{h}_{#1}^{\d}}

\providecommand*{\moire}{moir\'e}

\newcommand*{\heading}[1]{\belowpdfbookmark{#1}{#1}{\bfseries\textit{#1.---}}\ignorespaces}

\renewcommand*{\revised}[1]{\textcolor{red}{#1}}
\let\oldsout\sout
\renewcommand*{\sout}[1]{\revised{\oldsout{#1}}}

\renewcommand*{\revised}[1]{#1}

\newlength{\apsfigurewidth}
\global\long\def\k{\vec k}%
\newcommand*{\Q}{{\vec{K}}}
\global\long\def\p{\vec p}%
\global\long\def\R{\vec R}%
\global\long\def\K{\vec K}

\let\subsection\heading

\newcommand*{\citeapp}[1]{\cite{supplementary}}

\usepackage{comment}
\iffalse

\else
\excludecomment{gobble}
\fi

\everypar=\expandafter{\the\everypar\loosness=-1 }

\begin{document}
	
	%TC:ignore 
	\title{\revised{Leaky exciton condensates in transition metal dichalcogenide \moire{} bilayers}}
	
	\author{Benjamin Remez}
	\email{br395@cam.ac.uk}
	\affiliation{T.C.M. Group, Cavendish Laboratory, University of Cambridge, JJ Thomson Avenue, Cambridge CB3 0HE, United Kingdom}
	\author{Nigel R. Cooper}
	%\email[E-mail: ]{nrc25@cam.ac.uk}
	\affiliation{T.C.M. Group, Cavendish Laboratory, University of Cambridge, JJ Thomson Avenue, Cambridge CB3 0HE, United Kingdom}
	\affiliation{Department of Physics and Astronomy, University of Florence, Via G. Sansone 1, 50019 Sesto Fiorentino, Italy}

	\begin{abstract}	
		We show that the ``dark condensates" that arise when excitons form a Bose-Einstein condensate in a material with an indirect bandgap are not completely dark to optical emission.	Rather, such states are ``leaky condensates" in which optical emission is facilitated by many-body interactions. We analyze the properties of these leaky condensates in the context of twisted bilayers of transition metal dichalcogenides, which host strongly interacting excitons and an indirect bandgap. 
		\revised{We show that}  this interaction-driven ``leaky" emission \revised{dominates photoluminescence at low temperatures}, with distinctive qualitative features.
		\revised{Finally, we propose that in these materials, unique intervalley physics can lead to crystal symmetry-breaking excitonic ordering, with implications for optical processes.}
	\end{abstract}
	
	\date{\today}
	
	\maketitle
	
	%TC:endignore 
	
	Excitons, bound electron--hole (e--h) pairs, give rise to a plethora of quantum-coherent phenomena in solids, including light--matter hybridization \cite{Snoke2002, Byrnes2014}, long-range order \cite{Butov2002}, phase coherence \cite{High2012}, and Bose--Einstein condensates (BECs) \cite{MoskalenkoSnoke2000}. 
	Novel atomically-thin transition metal dichalcogenide (TMD) structures \cite{Tran2021}, featuring tightly-bound excitons with long lifetime \cite{Rivera2015,Palummo2015, Miller2017, Nagler2017, Jiang2018, Montblanch2021} and valley pseudospin with contrasting optical selection rules \cite{Jin2019b}, have spearheaded a new generation of excitonic devices.
	In parallel, the maturing field of twistronics \cite{Andrei2021} predicts phenomena such as flat \cite{Brem2020b} and topological excitonic bands with chiral edge modes \cite{Wu2017a}.
	This versatility is promising for realizing quantum emitters \cite{Yu2017, Baek2020}, simulators \cite{Kennes2021}, and exciton BECs \cite{Fogler2014, Berman2016}, and many-body exciton physics is being explored in  electrostatically gated, optically-inert excitonic insulators \cite{Jerome1967, Wu2015c, Debnath2017,Wang2019, Ma2021, Gu2021, Shi2021} and cavity exciton--polaritons \cite{Basov2016, Forg2019, Yu2020, Zhang2021a, Camacho-Guardian2021}. 
	
	In twisted TMD heterobilayers, interlayer excitons \cite{Rivera2018} formed by electrons and holes in opposite layers lie at low energies \cite{Wilson2017}, and provide a compelling platform for pumped exciton condensates. Firstly, the spatial separation of electrons and holes leads to long exciton lifetimes. The interlayer twist then rotates electron bands in momentum space \cite{Yu2015}, resulting in an indirect bandgap \cite{IntervalleyExcitons} and lifetimes longer still \cite{Rivera2015, Choi2021}. 
	Secondly, the misaligned layers form a large-scale \moire{} superlattice, with a spatially-modulated bandgap \cite{Zhang2017a,WuLovornMacDonald2018, Yu2017, Shabani2020, Guo2020, Brem2020a} that traps excitons in localized orbitals \cite{Tran2019, Mahdikhanysarvejahany2021, Karni2021, Jin2019, Seyler2019}.  This, and the excitons' interlayer electric dipole, place them in the strongly-interacting regime \cite{Yu2017}. These BECs thus merge strong correlations, quasi-equilibrium dynamics, opto-, twist-~and valleytronics. Clearly, new approaches are called for.

	In this Letter we show that the intersection of strong interactions and indirect gap leads to striking  optical properties in these \moire{} BECs. 
	The indirect bandgap suggests that the  exciton ground state forms a so-called ``dark condensate'' that cannot emit light directly \cite{Lagoin2021}.
	However, as we will show, no condensate is completely dark if interactions are considered\revised{. I}n this strongly-interacting system such effects are dominant\revised{, driving} emission from the BEC even at vanishing temperature, which we describe as a ``leaky condensate''. 
	These ``leaks'' give rise to  distinctive qualitative features in the optical emission of TMD \moire{} excitons at low temperatures.

	\begin{figure}[bt]
		\includegraphics[width=0.35\apsfigurewidth]{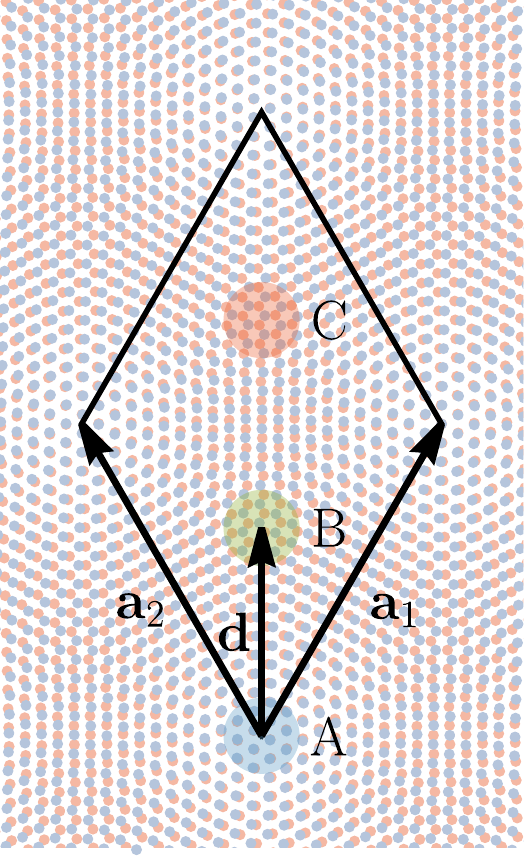}
		\includegraphics[width=0.64\apsfigurewidth]{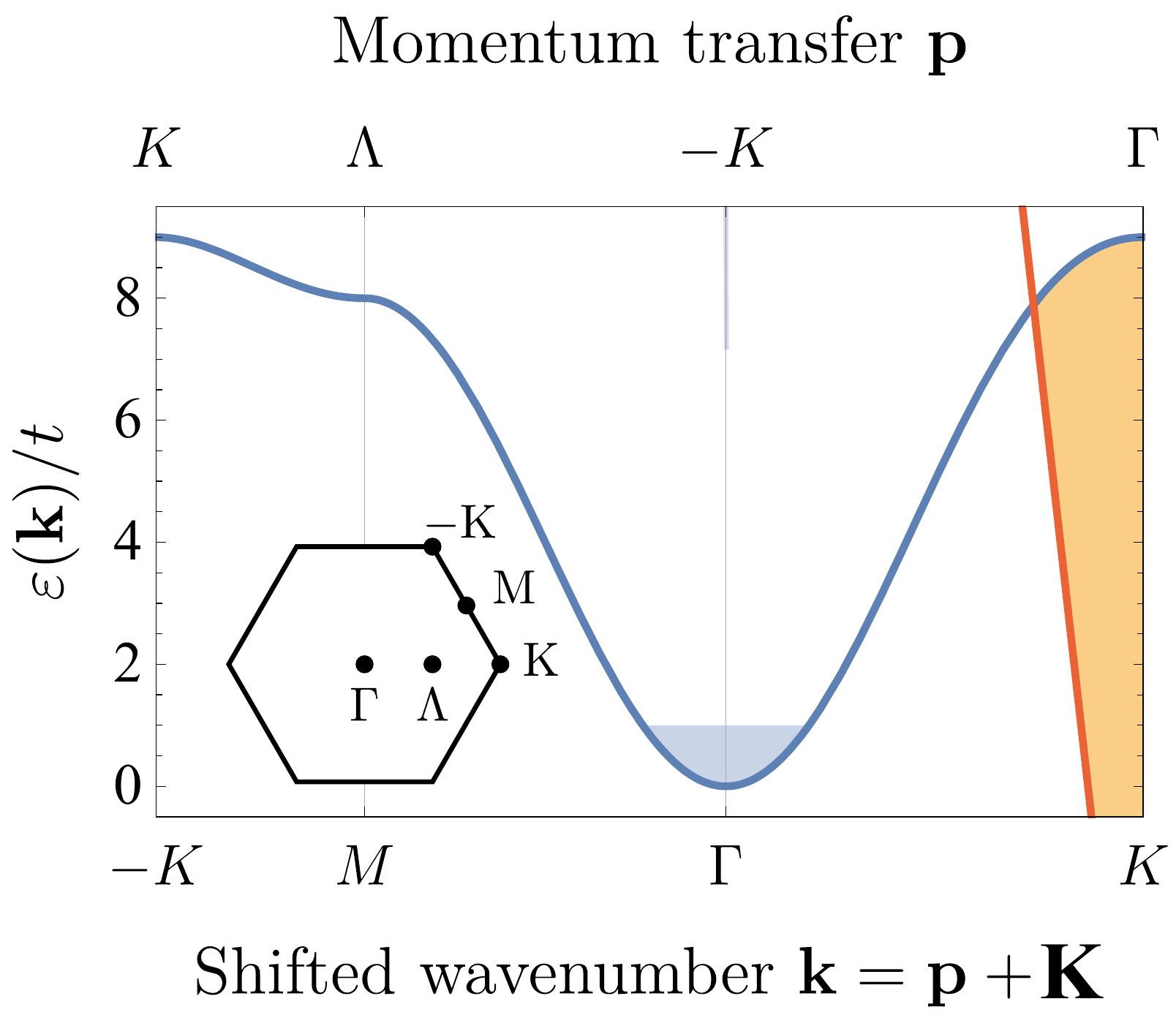}
		\caption{(a) The \moire{} triangular superlattice formed by a twist of $3^\circ$. The diamond outlines the \moire{} unit cell, with the three high-symmetry rotation centers A, B, and C highlighted in blue, green, and red. (b) The nearest-neighbor-hopping exciton dispersion. The blue and orange regions represent the condensate and the optical light cone (of $\tau = +1$ excitons, not to scale) respectively.}
		\label{fig:geometry}
	\end{figure}

	\subsection{Model}
	We consider a tight-binding lattice model for the interlayer excitons on the TMD \moire{} superlattice, as proposed in previous works \cite{Yu2017, WuLovornMacDonald2018, Jin2019, Seyler2019, Lagoin2021}. 	The superlattice inherits the triangular symmetry of the underlying monolayers, and has three high-symmetry locations labeled A, B, and C, as seen in Fig.~\ref{fig:geometry}a \cite{supplementary}. Sites A and B are local energy minima hosting bound states \revised{while C is a higher local energy maximum. A and B are generally not degenerate, and their energetic ordering may depend on the choice of monolayer compounds and whether they are stacked near 0 or 180 degrees} \cite{Yu2017, WuLovornMacDonald2018}. However, for simplicity we consider only the lowest-energy locale and our results hold whether it is A or B.
	\revised{Additionally, the triangular symmetry of each monolayer lends it a hexagonal Brillouin zone, with gapped valleys at the two inequivalent $\pm$K-point corners. 
	%Excitons are photogenerated by vertical interband transitions, and therefore can be composed of e--h pairs in either valley, which we label by the valley pseudospin index $\tau = \pm 1$ \cite{Jin2019b}
	e--h pairs are pumped optically via vertical interband transitions, and so excitons can be photogenerated in either valley, labelled by $\tau = \pm 1$ \cite{Jin2019b}.
	}
	 
	We  thus consider the Bose--Hubbard \cite{Fisher1989} Hamiltonian
	\begin{align} \label{eq:two-valley-Hamiltonian}
		\hat{H}  & 
		\!\!= \ns \ns \sum_{\k,\tau} \! (\!E_{0} \ns+\ns \epsilon_{\k}\!) \xdop{\k \tau\!} \xop{\k \tau\!} 
		\!+\!\!\ns\ns \sum_{\R, \tau,\sigma} \!\!\! \frac{U_{\!\tau \hspace{-0.075em} \sigma\!}}{2} \! \xdop{\R \tau\!} \xdop{\R \sigma\!} \xop{\R \sigma\!} \xop{\R \tau\!} 
		\!+\!  
		\hat{V}_\mathrm{LMI} \!\!\!\!
	\end{align}
	where $\xdop{\R \tau}$ is the bosonic creation operator of the lowest Wannier state in supercell $\R$ (at either locale A or B) \revised{and valley $\tau$}.
	$\xdop{\k \tau}$, defined shortly, creates plane-wave states with dispersion $\epsilon_\k$. We use $\epsilon_\k~=~-~t[4 \cos ( k_x a /2) \cos (\sqrt{3} k_y a /2) + 2 \cos(k_x a) - 6]$, corresponding to nearest-neighbor hopping with amplitude $t>0$ and \moire{} period $a$ \cite{supplementary}, and plotted in Fig.~\ref{fig:geometry}b.
	$E_0$ is the exciton formation energy (bandgap minus binding energy). $U_{\tau\sigma}>0$ are valley-dependent \cite{Rivera2016} on-site repulsion strengths; due to \moire{} localization a strongly-interacting system with $t/U_\mathrm{\tau\sigma} < 0.1$ is predicted \cite{Yu2017, Lagoin2021}, which we will treat accordingly.
	$\hat{V}_\mathrm{LMI}$ is the light--matter interaction, addressed below.
	
	A specific gauge is fixed in Eq.~\eqref{eq:two-valley-Hamiltonian}: 
	Exciton momentum eigenstates superpose e--h pairs with fixed momentum transfer $\p_e-\p_h=~\p$. Due to the indirect gap, the lowest-energy transition occurs at  $\p=-\tau\Q\neq\vec{0}$, where $\Q = \K_h - \K_e$ is the momentum mismatch between the $\tau=+1$ valley extrema of the two layers \cite{Yu2015, Yu2017, IntervalleyExcitons}, coinciding with the \moire{} Brillouin zone (MBZ) corner.
	It is  convenient to define the shifted wave vector $\k=\p+\tau\Q$, so real-~and $\k$-space states are related by
	\begin{equation} \label{eq:fourier_transform}
		\xop{{\k} \tau} = \frac{1}{\sqrt{N}} \sum_{\revised{\R}} e^{{-}i (\k - \tau\Q)\cdot \R} \xop {{\R} \tau}
	\end{equation} 
	with $N$ the number of supercells. Indeed, the hopping amplitudes $\braket{\R \tau | \hat{H} | \R' \tau}$ are complex. Their phases are fixed by the momentum mismatch \cite{Yu2017}, 
	which guarantees \citeapp{[cf. Appendix \ref{app:symmetries}]} that in this unique gauge $\epsilon_\k$ is valley-independent and minimal at $\k=\vec{0}$.  
	
	\revised{This momentum mismatch implies that the excitonic ground states cannot recombine radiatively: The nearly-vertical photon dispersion gives rise to the well-known optical light cone (LC) of states with $\left|\p\right| \apprle  E_0 / \hbar c$ that can recombine, with $c$ the speed of light in the surrounding medium. The light cone of each valley is thereby centered at $\k = \tau \Q$, i.e. the MBZ corners, see Fig.~\ref{fig:geometry}b. Crucially, the $\k=\vec{0}$ states lie outside both cones, and thus they are momentum-dark. }
	
	\revised{Under a broad set of conditions, the Hamiltonian \eqref{eq:two-valley-Hamiltonian} has a many-body ground state that is Bose-condensed, with a large phase-coherent occupation of the single-particle ground states at $\k=\vec{0}$. In direct-gap systems, where these states are bright, this would lead to a pronounced phase-coherent emission similar to superradiance \cite{MoskalenkoSnoke2000}. Yet the twist-induced momentum mismatch renders the BEC a ``dark condensate'' from which direct optical emission is forbidden by translation symmetry.}

	\revised{Breaking translation symmetry can enable emission from momentum-dark excitons. 
	This has been realized externally in TMDs, with electronic charge-order \cite{Shimazaki2021, Smolenski2021} and incommensurate substrates \cite{Joshi2021}. 
	%However, emission from the condensate requires a fine-tuned momentum folding between $\k=\vec{0}$ and $\Q$.
	Indeed, translation symmetry breaking is a hallmark of indirect-gap exciton coherence \cite{Jerome1967,Halperin1968,Kogar2017}. The interplay of two interacting valleys in Hamiltonian \eqref{eq:two-valley-Hamiltonian} can cause the excitons to break translation symmetry and form exciton density waves, which we explore in the Supplementary Material \citeapp{}. We find that while these density waves have precisely the required geometry, three-fold rotation symmetry prevents direct emission and the condensate remains dark.}

    \subsection{Leaky condensates} 
	While its coherent component cannot emit directly, the many-body condensed ground state is not completely dark. Rather, exciton--exciton interactions  induce an incoherent component that can radiate.  
	Emission is driven by excitonic collective modes \cite{Ozeri2005}, which have attracted recent attention as insightful probes of excitonic many-body states \cite{Utsunomiya2008,Kogar2017,Werdehausen2018ScienceAdvances,Remez2020, GoleZ2020, Estrecho2021,Bretscher2021}. Here, the collective modes supply the \revised{momentum}  necessary  for the excitons to recombine. 
	\revised{Collective modes exist in normal phases, and so Bose--coherence is not prerequisite for this  mechanism. However, below we demonstrate that in twisted bilayers this is the dominant emission channel at low temperatures. Thus, many-body interaction effects will determine the optical character of the low-temperature condensed phases, which we call ``leaky condensates''.}
	
	The essential physics of our mechanism is captured by a single-valley model. 
	This may be realized experimentally by pumping a valley-contrasting \revised{circularly-polarized} intralayer exciton resonance, followed by rapid interlayer charge transfer \cite{Zhang2019}.
	\revised{Excitons exhibit long valley depolarization times \cite{Jiang2018,Seyler2019, Zhang2019, Scuri2020} thanks to the large e--h vertical separation \cite{Rivera2018}, suggesting the exciton population can be treated as valley-polarized over radiative timescales.}
	%We thus project $\hat{H}$ 
	Projecting onto  valley $\tau=+1$ and suppressing $\tau$ henceforth, $\hat{H}$ reduces to% Eq.~\eqref{eq:two-valley-Hamiltonian} reduces to 
	\begin{equation} \label{eq:single_valley_hamiltonian} 
		\hat{H} = \!\sum_{\k} \! \left(E_0 + \epsilon_\k \right)\! \xdop{\k} \xop{\k} + \frac{U}{2}  \! \sum_{\R} \! \xdop{\R} \xdop{\R} \xop{\R} \xop{\R}+\hat{V}_\mathrm{LMI}. \!\!
	\end{equation}
	Eq.~\eqref{eq:fourier_transform} notwithstanding, all excitons now carry the same momentum mismatch so it is gauge-eliminable, and Eq.~\eqref{eq:single_valley_hamiltonian} realizes the usual Bose--Hubbard Hamiltonian.
	
	We model the light--matter interaction by 
	\begin{align} \label{eq:V_LMI}
		\hat{V}_\mathrm{LMI} = \!\!\!\sum_{\p_\parallel, p_\perp, \sigma}\!\!\! \hbar \omega_\p \adop{\p \sigma} \aop{\p \sigma} + (g_{\p\sigma} \adop{\p \sigma}\xop{\Q + \p_\parallel} + \hc),  
	\end{align}
	where $\adop{\p\revised{\sigma}}$ creates a photon with momentum $\p$ with the indicated in-~and out-of-plane components and polarization $\sigma$.
	We have incorporated the exciton momentum mismatch, and made the rotating-wave approximation \cite{ScullyZubairy1997}. The coupling constants $g_{\p\sigma}$ are obtained from electronic interband transition dipole matrix elements, and depend on the exciton pairing wavefunction \cite{Choi2021} etc. The recombination rate $\Gamma_\k$ of each $\k$ mode inside the light cone can be computed \cite{Choi2021} using Fermi's Golden Rule \cite{Sakurai2017}. This defines a natural timescale, the lifetime of a localized exciton in a single \moire{} site, 
	$\tau^{-1}_\mathrm{loc} = \tfrac{1}{N} \sum_{\k \in \mathrm{LC}} \Gamma_\k $. \revised{The relative size of the light cone compared to the total MBZ $=\tfrac{1}{N} \sum_{\k \in \mathrm{LC}} \sim (E_0 a / \hbar c)^2$  gives the fraction of the localized wavefunction that is contained in the light cone, and provides the estimate $\tau_\mathrm{loc} \sim (\hbar c / E_0 a )^2 / \Gamma_\Q \sim 10~\mathrm{ns}$  \cite{Yu2017}}. 
	Note $\tau_\mathrm{loc}$ is not the exciton mean radiative lifetime, which is associated with a thermal averaging over $\Gamma_\k$ \cite{Choi2021}.

	Let us assume that exciton recombination is sufficiently slow to maintain quasi-equilibrium, with an associated chemical potential $\mu$ and grand-canonical potential $\hat{\Xi}~=~\hat{H} - \mu \hat{\mathcal{N}}$. 
	We will study the ground state of the excitonic sector of $\hat{\Xi}$ and treat $\hat{V}_\mathrm{LMI}$ as a weak perturbation that generates photons which probe it. This picture is made consistent by shifting the photon energies to $(\hbar \omega_\p - \mu)$ \citeapp{}.

	\subsection{Weak interactions: Bogoliubov theory}
	Though the excitons we consider are strongly-interacting \cite{Yu2017, Lagoin2021}, we first present our emission mechanism in the  familiar weakly-interacting Bogoliubov theory to build our intuition. 
	
	A BEC with all excitons at $\k=\vec{0}$  will be depleted by interactions that  eject pairs of counter-propagating excitons from the condensate. 
	If one lands within the light cone, it may recombine.  
	Bogoliubov theory lets us neatly resum these virtual processes and find the depleted ground state. 
	
	Consider a state with total filling $\nu$ and condensate filling $\nu_c(\nu)$.
	Eq.~\eqref{eq:single_valley_hamiltonian} leads to the standard \cite{BECPitaevskiiStringari} mean-field (MF) Bogoliubov--de-Gennes (BdG)  Hamiltonian
	$\hat{\Xi}_\mathrm{MF} = \sum_\k  \Omega_\k \bdop{\k}\bop{\k}$
	with the familiar dispersion $ \Omega_\k^2 = {\epsilon_\k (\epsilon_\k + 2 \nu_c U)}$, $\mu=E_0 + \nu_c U$, and the Bogoliubov modes 
	\begin{align} \label{eq:bogoliubov_modes}
		\bop{\k} & = \cosh(\theta_\k) \xop{\k} \!+\! \sinh (\theta _\k) \xdop{-\k}, \,\,\,
		\sinh \theta_\k 
		= \frac{\Omega_\k \!-\! \epsilon_\k}{2 \sqrt{ \Omega_\k \epsilon_\k}}.\!\!
	\end{align}
	The ground state of the theory is the BdG vacuum. Yet the mixing of particle creation and annihilation in Eq.~\eqref{eq:bogoliubov_modes} implies that it nevertheless contains some excitons, most notably inside the light cone.
	
	Consider recombination in terms of the collective modes. Transcribed into BdG modes, Eq.~\eqref{eq:V_LMI} contains terms such as
	$- g_{\p\sigma}  \sinh (\theta_{\Q + \p_\parallel}\!) \adop{\p \sigma} \bdop{-\Q - \p_\parallel}$, which
	represent a spontaneous emission of a photon and a BdG mode, the latter assuring momentum conservation in analogy to phonon-assisted exciton recombination \cite{MoskalenkoSnoke2000}. Therefore, interactions enable an otherwise-dark exciton condensate to ``leak'' photons with 
	\begin{equation} \label{eq:leaky_emission_spectrum}
		\hbar \omega_\p  = \mu - \Omega_{-\Q - \p_\parallel} = E_0 + \nu_c U -  \Omega_{-\Q - \p_\parallel} < E_0\,.
	\end{equation} 
	As expected, some energy is lost to the Bogoliubov mode.
	
	We  compute the total  emission rate $\Gamma$ with Fermi's Golden Rule via transitions between the BdG vacuum and single quasiparticle states. The redshift in Eq.~\eqref{eq:leaky_emission_spectrum} is negligible compared to $E_0$,  
	so this is equivalent to counting the number of excitons  present within the light cone, 
	\begin{align} \label{eq:leaky_emission_rate} 
		\Gamma   & 
		\approx \sum_{\k \in \mathrm{LC}} \Gamma_\k n_\k 
		\approx N \tau^{-1}_\mathrm{loc} \, n_\Q 
		\approx 
		N \tau^{-1}_\mathrm{loc}  \Big(\dfrac{1}{18}  \dfrac{\nu_c}{t/U}\Big)^2. 
	\end{align}
	Here we approximated $n_{\k} = \braket{\xdop{\k}\xop{\k}}\approx n_\Q =
	\sinh^2 \theta_{\Q}$ inside the light cone, assuming it is much smaller than the MBZ, and expanded around small densities. The total filling is found by integrating $n_\k$ \cite{BECPitaevskiiStringari}, and in two dimensions $\nu_c = \nu [1 - \mathcal{O}(U/t)]$. Thus, unlike spontaneous decay, $\Gamma$ is quadratic in density.

	The generality of this construction suggests  that exciton condensates in any indirect-gap system are leaky. Moreover,  emission is accomplished without additional degrees of freedom. This contrasts with external optical probing \cite{Combescot2014a} and other mechanisms that  involve phonons \cite{Danovich2016} or carrier exchange in larger exciton complexes \cite{Danovich2017}.
	
	Interactions  deplete excitons into excited bands as well, and so leaky condensates can also occur in systems that are dark due to a spin-forbidden transition, etc. Previous studies have shown that, for sufficiently strong interactions (or, equivalently, above a threshold density), a dark condensate can  transition into a  so-called ``gray condensate'' \cite{Combescot2012, Alloing2014, Combescot2017, MazuzHarpaz2019}. Our finding of leaky emissions is distinct from these previous works: 
	(i)~Emission from the leaky condensate grows continuously with increasing density,  with no threshold value.
	\revised{(ii)~In indirect-gap materials, the bright and (momentum-)dark states are smoothly connected along the same Bloch band, unlike the usual scenario where they are spin-split and form separate bands. This prevents fragmentation into a gray condensate.}
	(iii)~The gray condensate emission is coherent whereas the leaky condensate emission is not, due to its entanglement with the generated collective modes.

	Condensate depletion by interactions has recently gained significant experimental attention \cite{Xu2006, Chang2016,Lopes2017, Pieczarka2020, Steger2021}. 
	Relatedly, BdG modes are used to renormalize phonon-assisted photoluminescence line shapes, e.g. in Cu$_2$O \cite{MoskalenkoSnoke2000, Haug1983, Shi1994}. However, the role of collective modes in {\emph{enabling}} recombination of indirect-gap excitons, to the best of our knowledge, has not been pointed out so far.
	Additionally, these descriptions focus on weakly-interacting excitons.
	
	\subsection{Strong interactions: hard-core bosons}
	We  now explore how  leaky condensates manifest under strong interactions. Consider the $U \to \infty$ limit corresponding to hard-core bosons, which with the transformation ($S=\frac{1}{2}$ henceforth implied) 
	\begin{equation} \label{eq:spin_operators}
		\sop{\R}^- = e^{-i\Q\cdot\R} \xop{\R},\,\,
		\sop{\R}^+ = e^{i\Q\cdot\R} \xdop{\R},\,\,
		\sop{\R}^z= \xdop{\R} \xop{\R} \!-\! S, 
	\end{equation}
	map to a spin-$\frac{1}{2}$ XX ferromagnet in a transverse field. The grand-canonical potential becomes 
	\begin{align} \label{eq:XXZ-Hamiltonian}
		\hat{\Xi} & 
		\!=\! -t \!\!\!\! \sum_{\langle \R,\R'\rangle} \!\!\!\!  ( \sop{\R'}^+ \sop{\R}^- \!+\! 
		\sop{\R}^+ \sop{\R'}^- ) 
		+ (E_0 \!-\!\mu ) \! \sum_{\R} \! (\sop{\R}^z \!+\! S ).  
	\end{align}
	The recombination rate remains $\Gamma\tau_\mathrm{loc} /N~\approx n_{\Q}~=~\braket{\sop{\Q}^+ \sop{\Q}^-}$ where $\sop{\Q}^-~=~\frac{1}{\sqrt{N}} \sum_\R e^{-i\Q\cdot\R} \sop{\R}^-$. 
	
	This limit readily manifests ground state emission: With $t = 0$, Eq.~\eqref{eq:XXZ-Hamiltonian} factorizes into independent sites each with MF solution $\sqrt{\nu}\ket{\Uparrow} + \sqrt{1-\nu} \ket{\Downarrow}$, yielding \citeapp{} $n_{\Q}(\nu)=\nu^2$. As nonzero hopping allows repelling excitons to separate, leading to \emph{anti}correlations, this is an upper bound. Additionally, an emergent particle--vacancy duality \citeapp{} $\sop{\R}^\pm \to \sop{\R}^\mp$ connects the ground states of Eq.~\eqref{eq:XXZ-Hamiltonian} with fillings $\nu$ and $1-\nu$, providing the identity $n_\Q (\nu) - n_\Q (1-\nu) = 2\nu - 1$. Thus, $n_\Q(\nu)>\mathrm{max}(2\nu -1,0)$. These bounds already demonstrate the interaction-driven nonlinearity of $\Gamma (\nu)$. 
	
	A quantitative treatment of emission is again found in terms of  spontaneously excited collective modes, now taking the form of spin waves. We perform a Holstein-Primakoff (HP) $1/S$ expansion \cite{HolsteinPrimakoff1940} similar to Bernardet et~al.~\cite{Bernardet2002} with details provided in the Supplemental Material \citeapp{}. 
	The qualitative features of Bogoliubov theory are reproduced: we obtain a quadratic  $\hat{\Xi}=\sum\Omega_\k \bdop{\k}\bop{\k}$, now with dispersion $\Omega_\k^2 = \epsilon_\k [(2\nu_\mathrm{MF}-1)^2\epsilon_\k + 24\nu_\mathrm{MF}(1-\nu_\mathrm{MF}) t] $ and $\mu = E_0 + 6(2\nu_\mathrm{MF}-1)$, where the filling $\nu_\mathrm{MF}$ is the mean-field order parameter [cf. $\nu_c$ in the previous section].
    The exciton occupation \revised{in the spin-wave vacuum, or equivalently at temperature $T=0$, is now} 
    \begin{equation}
        n_\k % = \braket{\xdop{\k}\xop{\k}} 
        = \left[\nu_\mathrm{MF} \cosh \theta_\k + (1-\nu_\mathrm{MF}) \sinh \theta_\k\right]^2.
    \end{equation}
	 $\theta_\k$ and the emission spectra are still given by Eqs.~\eqref{eq:bogoliubov_modes} and \eqref{eq:leaky_emission_spectrum}.

	\begin{figure}[tb]
		\includegraphics[width=\apsfigurewidth]{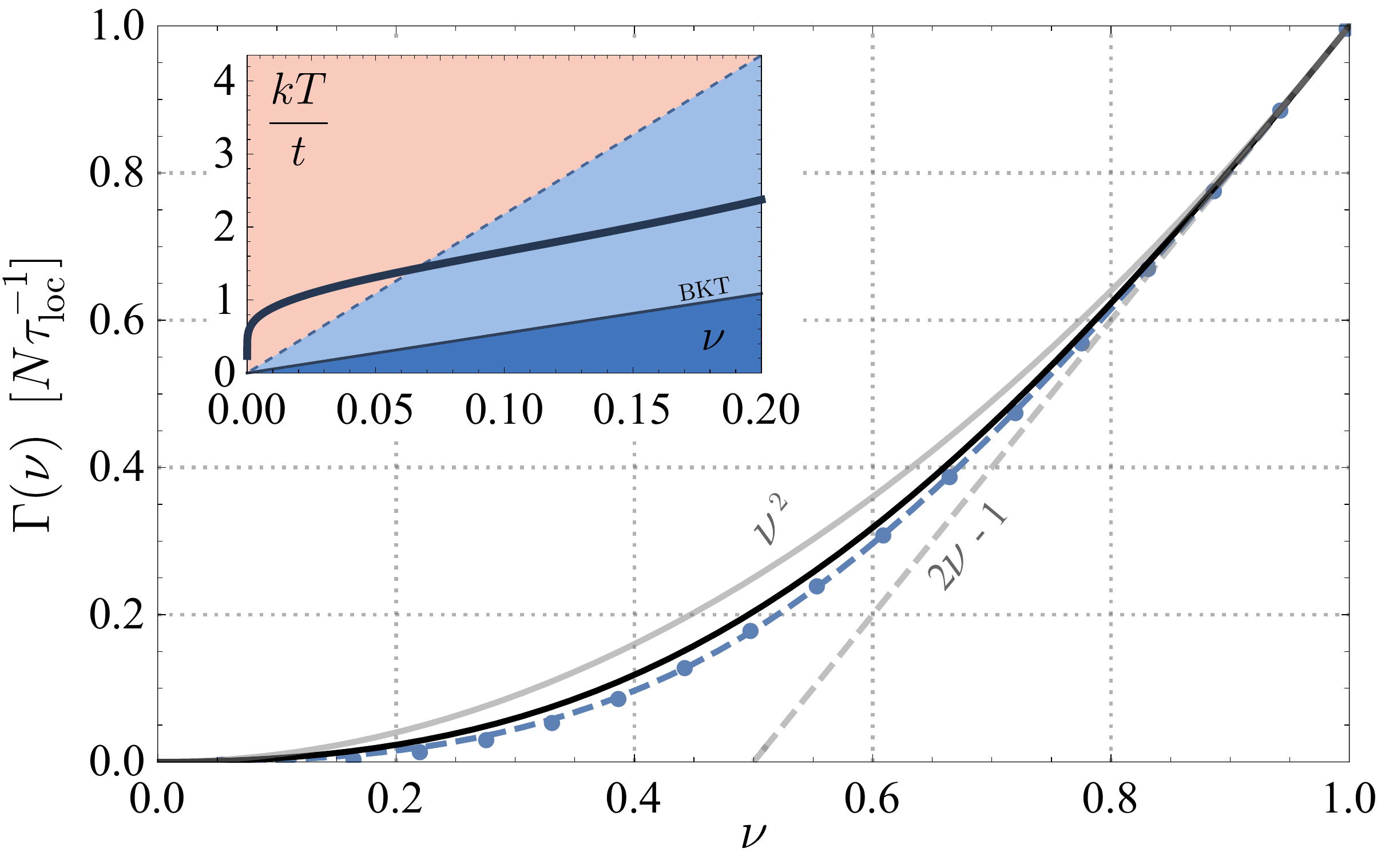}
		\caption{Emission rate $\Gamma(\nu)$ \revised{at $T=0$} versus filling \cite{densities} within the Holstein--Primakoff expansion \citeapp{} (solid black), compared with exact diagonalization on a 6$\times$3 lattice (filled circles) and a least-squares fit derived from two-body states (dashed blue). 
			The solid and dashed gray lines are upper and lower bounds explained in the text.
			Inset: The crossover temperature $T^\star$ between interaction- and thermally-dominated emission. 
			The solid and dashed thin lines mark the Berezinskii–Kosterlitz–Thouless superfluid transition and quantum degeneracy crossover, estimated at $kT/t=\sqrt{3}\pi\nu$ and $4\sqrt{3}\pi\nu$, respectively \citeapp{}. Interactions dominate the  cold regimes.}
		\label{fig:hard-core_boson}
	\end{figure}

	We plot the \revised{$T=0$} emission rate $\Gamma\propto n_\Q$ in Fig.~\ref{fig:hard-core_boson} \cite{densities}. Comparing it against small-scale exact diagonalization, we find very good agreement across a wide range of fillings. 
	In the dilute limit
	\begin{equation} \label{eq:hard-core_bosons_emission_rate}
		\Gamma(\nu \ll 1) \approx  
		\frac{4}{9}  N \tau_\mathrm{loc}^{-1} \nu_\mathrm{MF}^2 .
	\end{equation}
	The apparent quadratic dependence in Eq.~\eqref{eq:hard-core_bosons_emission_rate} does not imply an absence of correlations, which  cause
	$\nu_\mathrm{MF}(\nu)$ and $\nu$ to differ. Exciton correlations can be inferred from two-body states \cite{MazuzHarpaz2019}, from which we deduce \citeapp{} the asymptotic form $\Gamma \sim (\nu/\log \nu)^2$, revealing the expected suppression. 
	Numerics confirm \citeapp{} that $\nu_\mathrm{MF}\sim(\nu/\log \nu)$ for $\nu\ll1$, indicating correlations are successfully captured.

	Our HP theory also allows us to treat nonzero temperatures. Thermal excitations enhance emission and unlock a second channel whereby a collective mode is absorbed instead of emitted, leading to two emission lines. 
	In the inset of Fig.~\ref{fig:hard-core_boson} we plot the crossover temperature $T^\star$ at which depletion (due to interactions) and thermal excitations contribute equally \citeapp{} to the total emission rate, and below which interactions dominate. 
	This crossover occurs above the Berezinskii--Kosterlitz--Thouless (BKT) transition \cite{Berezinskii1972, KosterlitzThouless1973}, indicating that \revised{leaky emission will be the dominant emission channel characterizing the superfluid phase and much of the quantum-degenerate regime}.
	\revised{Remarkably, in this strongly-interacting system, at small filling ``leaks'' dominate emission even in the hot gas phase that can be treated semi-classically.} 
	Unlike the  Stokes and anti-Stokes lines in phonon-assisted emission \cite{MoskalenkoSnoke2000}, the two BdG processes have unequal matrix elements with different density dependences. The anti-Stokes-like line dominates above $T^\star$, and the net annihilation of BdG modes may evaporatively cool the BEC,  similarly to a mechanism recently suggested \cite{Misra2021}.
	Finally,  the strength of interactions can be inferred from the ratio of emission line intensities \citeapp{}.

	The leaky condensate picture thus predicts a distinctive property of \moire{} exciton emission: A dominant redshifted emission line with  quadratic density dependence below $T^\star$, compared to a dominant blueshifted line with linear density dependence above $T^\star$.
	We do not expect a qualitative change at the BKT transition.
	
	\subsection{Experimental Consequences}
	We assess the parameters under which  a leaky condensate may be observed.
	Comparing Eq.~\eqref{eq:leaky_emission_rate} with \eqref{eq:hard-core_bosons_emission_rate} suggests that this mechanism saturates once $U \apprge 12 t$, which should   hold across a wide range of twist angles \cite{Yu2017}.
	For  $a \approx 10~\mathrm{nm}$ (twist $\approx 2^\circ$)  and corresponding $t \approx 0.2~\mathrm{meV}$ \cite{Yu2017}, and at a demonstrated \cite{Wang2018a} photoexcited exciton densities of $n \approx 10^{11}~\mathrm{cm}^{-2}$ ($\nu \approx 0.1$),  $T^\star \approx 5~\mathrm{K}$. 
	Larger twist angles or intercalated hBN spacers would increase $t$ and thus $T^\star$ \cite{Yu2017, Lagoin2021}, and mitigate inhomogeneity effects. Furthermore, electrostatic gating might modify the \moire{} symmetry \cite{Yu2017} or its elastic reconstruction \cite{Andersen2021, BennettRemez2021}, allowing in-situ tunability.
	Above $T^\star$ the two emission lines are split by $\sim 20t\sim 4~\mathrm{meV}$ and should be resolvable.

	Leaky emission will also manifest in a quadratic loss $\partial_t n = - \gamma n^2 $, with $n$ the exciton number density. With Eq.~\eqref{eq:hard-core_bosons_emission_rate} and $\tau_\mathrm{loc}\sim10~\mathrm{ns}$ at $a\sim10~\mathrm{nm}$ \cite{Yu2017} we estimate a rate constant $\gamma\sim a^2/\tau_\mathrm{loc} \approx 10^{-4}~ \mathrm{cm^2/s}$.
	Exciton density is controlled with pumping fluence, and tracked with transient absorption \cite{Sun2014, Poellmann2015}, emission blueshift [cf.~Eq.~\eqref{eq:leaky_emission_spectrum}] \cite{Alloing2014,Combescot2014a,Wang2018a}, or time-resolved photoluminescence (PL) \cite{Wang2018a}. These reveal such quadratic dependence, which is attributed to Auger recombination. In TMDs this process is associated with  \revised{a high-energy conduction band that supports bound excitons at energies close to $2 E_0$, making Auger recombination nearly resonant} \cite{Steinhoff2021, Lin2021}. Yet crucially, for nonzero detuning Auger recombination freezes out at $T\to0$ \cite{Erkensten2021} and leaky emission will dominate. Comparisons at nonzero temperature are not straightforward; estimates are available mostly for room-temperature monolayers,  suggesting an intrinsic Auger constant $\gamma \sim 10^{-3}~\mathrm{cm^2/s}$ \cite{Hoshi2017, Zipfel2020, Steinhoff2021}. However, moving to cryogenic temperatures or to bilayers could each reduce $\gamma$ by orders of magnitude \cite{Erkensten2021,Yuan2015}. Thus, the dominance of leaky recombination over Auger could feasibly extend to $T^\star$ and above. 
	This unique regime of emission linear in continuous fluence yet nonlinear in \emph{instantaneous} density might be important in interpreting PL experiments. In settings demonstrating population lifetimes approaching microseconds \cite{Miller2017, Nagler2017, Jiang2018, Montblanch2021}, the short "leaky lifetime" $\tau_\mathrm{leaks}\sim (\gamma n)^{-1} \sim 10$ ns induced at high fluences could also play a significant role in the complex population dynamics.

	\subsection{Conclusions}
	We have shown that strong interactions challenge the picture of dark  condensates in \moire{} bilayers, where they dominate optical processes. This prompts further material-specific modelling to compare these effects to other recombination and loss mechanisms.
	\revised{Additionally, while here we mostly considered a single-valley model, the physics of two-valley \moire{} condensates is very rich. As we remark above, excitonic density waves put such condensates on the verge of optical activation which is prevented only by vestigial rotational symmetry \citeapp{}. Therefore, direct emission with long-range phase coherence may be achieved in these systems by external fields, strain, layer separation, and pressure, with possible sensing applications. We leave this and other novel intervalley phenomena to future work.}
	
	%TC:ignore 
	
	\subsection{Acknowledgments}
	We thank M.~Atat\"ure, D.~Kara, C.~M.~Pursar, A.~R.-P.~Montblanch,  T.~F.~Heinz, O.~Karni, E.~Barr\'e, A.~Camacho-Guardian and D.~Bennett for fruitful discussions.  The support of the Cambridge International Trust,  of EPSRC Grant Nos. EP/P009565/1, EP/P034616/1 and of a Simons Investigator Award are gratefully acknowledged.

	\bibliographystyle{apsrev4-1}
	\bibliography{references.bib,auxilary_refs.bib}

\begin{thebibliography}{110}
\bibitem [{den()}]{densities}%
  \BibitemOpen
  \href@noop {} {}\bibinfo {note} {The Holstein-Primakoff approximation is
  generally non-conserving. To recover the particle--vacancy duality of
  Eq.~\eqref{eq:XXZ-Hamiltonian}, in Fig.~\ref{fig:hard-core_boson} we plot
  $\Gamma(\nu)$ versus $\nu_\mathrm{MF}$ \citeapp{}.}\BibitemShut {Stop}%
\end{thebibliography}
	
	\clearpage

	%%%%%%%%%%%%%%%%%%%%%%%%%%%%%%%%55
	%  SUPPLEMENTARY
	%%%%%%%%%%%%%%%%%%%%%%%%%%%%%%%%%
	
	\pagebreak
	\begin{widetext}
	\begin{center}
		\textbf{\large Supplemental Material: Leaky exciton condensates in transition metal dichalcogenide \moire{} bilayers} \\
		\vspace{0.25cm}
		{Benjamin Remez$^1$, Nigel R. Cooper$^{1,2}$}\\
		\vspace{0.25cm}
		\textit{\small ${}^1$ T.C.M. Group, Cavendish Laboratory, University of Cambridge, JJ Thomson Avenue, Cambridge CB3 0HE, United Kingdom \\
		${}^2$ Department of Physics and Astronomy, University of Florence, Via G. Sansone 1, 50019 Sesto Fiorentino, Italy}
		\vspace{0.25cm}
	\end{center}
	\end{widetext}
	%%%%%%%%%% Merge with supplemental materials %%%%%%%%%%
	%%%%%%%%%% Prefix a "S" to all equations, figures, tables and reset the counter %%%%%%%%%%
	\setcounter{equation}{0}
	\setcounter{figure}{0}
	\setcounter{table}{0}
	\setcounter{page}{1}
	\makeatletter
	\renewcommand{\theequation}{S\arabic{equation}}
	\renewcommand{\thefigure}{S\arabic{figure}}
	%\renewcommand{\bibnumfmt}[1]{[S#1]}
	%\renewcommand{\citenumfont}[1]{S#1}
	%%%%%%%%%% Prefix a "S" to all equations, figures, tables and reset the counter %%%%%%%%%%

		\section*{Preliminaries} \label{app:symmetries}
	
	In this Section we briefly outline the basic parameters of the \moire{} lattice used  throughout the main text.
	
	\subsection{Geometry}
	We define the lattice primitive vectors
	\begin{equation}
		\vec{a}_1 = a\hat{\vec{x}}, \qquad 
		\vec{a}_2 = \frac{1}{2}a\hat{\vec{x}} + \frac{\sqrt{3}}{2} a \hat{\vec{y}},
	\end{equation}
	where $a$ is the \moire{} period. The \moire{} lattice has symmetry group p3m1, with 3 high-symmetry points A, B, and C. These locales are found at relative displacements
	\begin{equation}
		\vec{d}_\mathrm{A} = \vec{0}, \quad \vec{d}_\mathrm{B} = \vec{d}, \quad \vec{d}_\mathrm{C} = 2\vec{d},\quad \vec{d} = (\vec{a}_1 + \vec{a}_2) / 3.
	\end{equation}
	within each unit cell.	
	The reciprocal basis vectors are
	\begin{equation}
		\vec{b}_1 = \frac{2\pi}{a}(\hat{\vec{x}} -\frac{1}{\sqrt{3}}\hat{\vec{y}}), \qquad
		\vec{b}_2 = \frac{2\pi}{a} \frac{2}{\sqrt{3}} \hat{\vec{y}}.
	\end{equation}
	
	The \moire{} Brillouin zone has two inequivalent corners which should be labeled consistently.
	We follow Refs.~\cite{WuLovornMacDonald2018, Yu2017} which find the A and B sites host s-wave bound states with angular momenta $\ell_\mathrm{A} = \tau$ and $\ell_\mathrm{B} = -\tau$, respectively. 
	The C locale is a potential maximum and so does not host any states. \revised{The three-fold ${C}_\mathrm{3}$ rotation symmetry of the bilayer (see below)} dictates that the hopping amplitudes from an A site to its three nearest-neighbor B sites must be complex and transform like a $d_{+2}$ wave (mod 3),
	\begin{equation} \label{eq:hopping_parameters_d_wave}
		t(C_3^2\vec{d}) = e^{-\frac{2\pi i}{3} \tau} t(C_3 \vec{d}) =  e^{-\frac{4\pi i}{3} \tau} t(\vec{d}),
	\end{equation}
	where $t(\vec{v}) = \braket{\R + \vec{v} , \tau | \hat{H} | \R , \tau } $. On the other hand, the (intravalley) hopping amplitude phases are fixed by momentum mismatch \cite{Yu2017}
	\begin{equation}   \label{eq:hopping_parameters}
		t(\vec{v}) =  - \left|t(\vec{v})\right| e^{-i\tau\Q \cdot \vec{v}},
	\end{equation} 
	where $(-\tau\Q)$ is the exciton momentum mismatch, $\left|t(\vec{v})\right|$ has three-fold symmetry, and we separated the usual minus sign. $\Q$ coincides with one of the \moire{} Brillouin zone corners, and Eqs.~\eqref{eq:hopping_parameters_d_wave} and \eqref{eq:hopping_parameters} are reconciled by picking 
	\begin{equation}
		\Q = \frac{2}{3} \vec{b}_1 + \frac{1}{3} \vec{b}_2 = \frac{4 \pi}{3 a} \hat{\vec{x}}.
	\end{equation}
	
	\subsection{Kinetic energy} 
	The valley-dependent complex phases \eqref{eq:hopping_parameters} generally lead to valley-dependent dispersions. However, the gauge choice \eqref{eq:fourier_transform} uniquely gives a valley-independent dispersion, 
	\begin{align}
		\epsilon_\k & = 
		\braket{\k,\tau|\hat{H}|\k,\tau}  
		= \frac{1}{N} \! \sum_{\R, \vec{v}} e^{-i(\k-\tau\Q)\vec{v}} \! \braket{\R + \vec{v}, \tau | \hat{H} | \R, \tau} \tcbr
		= -\sum_\vec{v} \left|t(\vec{v})\right| e^{-i\k \cdot \vec{v}}.
	\end{align}
	We see this gauge also conveniently fixes the minimum of $\epsilon_\k$ to $\k=\vec{0}$. In this work we consider only the lowest energy locale, and hopping to the same locale in adjacent supercells. We thus set $\left|t(C^n_6 \vec{a}_1)\right| = t$ and otherwise zero. In the main text and here throughout we absorb $\epsilon_{\vec{0}}$ into $E_0$ so that the band minimum is at 0 and $\epsilon_\Q = 9t$.

	\section*{Exciton density waves \& selection rules}

	In the main text we explore the single-valley model of Eq.~\eqref{eq:single_valley_hamiltonian}, and direct emission from the BEC is forbidden by translation symmetry since the $\k=0$ excitons are momentum-dark.
	However, if translation symmetry is broken, recombination of these excitons will be permitted. 
	Such brightening of large-momentum excitons has been demonstrated in TMD mono- and bilayers via doped electronic charge-ordered states or interfacing with an incommensurate substrate \cite{Shimazaki2021, Smolenski2021, Joshi2021}.
	Motivated by these studies, we point out that the interplay between the two valleys in the full Hamiltonian of Eq.~\eqref{eq:two-valley-Hamiltonian} gives rise to excitonic density order which reduces translation symmetry. 
	However, emission from the condensate requires a fine-tuned momentum transfer between $\k=\vec{0}$ and $\Q$. 
	To determine whether the BEC becomes bright thanks to this density wave, or if it remains dark, we analyze the symmetries of the full $\hat{H}$, and  those that might be spontaneously broken by a two-valley condensate.
	
	\subsection{Translation symmetry}
	Translation symmetry breaking is a paradigmatic feature of indirect-gap exciton coherence \cite{Jerome1967,Halperin1968,Kogar2017}. 
	Here, a BEC comprised of excitons in both $\tau=\pm1$ valley minima, having different momenta $\p=\mp \Q$, is no longer an eigenstate of total momentum $\hat{\vec{\mathcal{P}}}$, showing translation symmetry is lost and momentum is not conserved. This is understood as an interference between the $e^{\mp i \Q\cdot\R}$ components, leading to a spatially-moduled exciton density wave. 
	
	However, translation symmetry is not broken completely, and instead reduced to that of the larger density wave periodicity. This implies a zone-folding scheme rather than mixing between any two arbitrary momenta, and the density wave might not necessarily couple the BEC and the light cone. Crucially, in our case the necessary momentum-folding is achieved not because the two valleys have opposite mismatch $\p = -\tau\K$ that sum to zero, but because their mismatch \emph{difference} equals the BEC--light cone offset $\Q$, since $\Delta \p = -2\Q=+\Q$ modulo \moire{} reciprocal lattice vectors. Therefore, a potential brightening of the BEC is also a special feature of the \moire{} geometry and valley configuration.
	
	Additionally, we note that appearance of the density wave also necessitates interactions, as integrability imposes a significantly stronger symmetry than the conservation laws we consider here. Therefore, like the leaky emission, the density wave is understood as a feature of the strong-interactions regime of interlayer excitons.

	It is thus instructive to demonstrate explicitly the origin of the density wave. Eq.~\eqref{eq:two-valley-Hamiltonian} contains, among others, cross-valley interaction terms like $\sum_{\R} \xdop{\R\tau}\xdop{\R\bar{\tau}}\xop{\R\bar{\tau}}\xop{\R\tau}$ with the notation $\bar{\tau}=-\tau$. The mean-field-decoupled Hamiltonian is obtained from the different ways $\hat{H}$ can be contracted. Recalling Eq.~\eqref{eq:fourier_transform}, upon condensation of the $\k=\vec{0}$ states, $\braket{\xop{\R\tau}}\propto e^{-i\tau\Q\cdot\R}$. Thus, in the presence of a two-valley condensate,  %$\braket{\xdop{\R\tau} \xop{\R \bar\tau}}\propto e^{2i\Q\cdot\R}$, and 
	the contraction  
	\begin{equation}
		\hat{H}_\mathrm{MF} \ni U_{\tau \bar{\tau}}\sum_\R \left(\braket{\xdop{\tau} \xop{ \bar\tau}} e^{2i\Q \cdot\R}  \xdop{\R \bar\tau}  \xop{\R \tau}+\hc\right) 	
	\end{equation}
	emerges.
	These $\pm2\Q$ Fourier components in the MF effective Hamiltonian demonstrate the translation-symmetry-breaking density wave. 
	Thanks to the \moire{} geometry, $2\Q = 3\Q-\Q = 2\vec{b}_1 + \vec{b}_2 - \Q=-\Q$ modulo the \moire{} Brillouin zone. Thus this term represents \emph{Umklapp} scattering of the excitons by $\pm\Q$. Consequently all the high-symmetry momenta states at $\p=0,\pm\Q$ are coupled together, showing the light cones are folded onto the condensate. 
	
	However, as we now explain, other symmetries of $\hat{H}$ will prevent emission even in this translation-symmetry-broken BEC. 
	
	\subsection{Intervalley symmetry}
	The long exciton valley depolarization time \cite{Jiang2018,Seyler2019, Zhang2019, Scuri2020} enabled by the large e--h separation \cite{Rivera2018} motivates an effective U(1) symmetry of separate particle conservation in each valley at short times, and Eq.~\eqref{eq:two-valley-Hamiltonian} does not contain valley-flipping terms.
	Such symmetry would prohibit genuine momentum folding,
	as it implies that
	\begin{equation}
		\hat{\vec{\mathcal{K}}}~=~\hat{\vec{\mathcal{P}}} + \Q(\hat{\mathcal{N}}_+ - \hat{\mathcal{N}}_-),
	\end{equation}
	with $\mathcal{N}_\tau$ the population in each valley, is a constant of motion with $[\hat{\vec{\mathcal{K}}}, \hat{H}]=\vec{0}$. Unlike  $\hat{\vec{\mathcal{P}}}$, $\hat{\vec{\mathcal{K}}}$ remains conserved by the density wave (i.e. commutes with $\hat{H}_\mathrm{MF}$) since both valley minima are its $\k = \vec{0}$ eigenstates.
	Thus modes of differing $\k$ cannot mix and momentum remains unfolded.
	
	We then consider a valley interconversion perturbation
	\begin{equation}
		\Delta\hat{H}_\mathrm{IV}~=~\sum_{\k,\tau} J^\mathrm{IV}_{\k \tau} \xdop{\k + \bar{\tau}\Q, \bar{\tau}} \xop{\k + \tau\Q, \tau}
	\end{equation}
	with intervalley tunneling coefficients  $J^\mathrm{IV}_{\k \tau}$. This is the exchange interaction that leads to longitudinal--transverse splitting of exciton modes in TMD monolayers \cite{Yao_LT_splitting}, which should also exist for interlayer excitons, albeit with reduced size due to their spatially indirect character \cite{Seyler2019}. Using   $J^\mathrm{IV}_{\k \tau}$ given by nearest-neighbor hopping commensurate with the crystal symmetries, we confirm by an explicit mean-field theory calculation that the two-valley BEC indeed leads to a translation-symmetry-broken effective Hamiltonian. However, despite being folded over each other, the condensate and light cones remain decoupled.
	
	\begin{table}[tb]
		\hspace*{\fill}
		%\begin{ruledtabular}
		\begin{tabular}{ccccccccc} \toprule \toprule
			\multicolumn{1}{c}{host} & \multicolumn{1}{c}{} & \multicolumn{3}{c}{$(\k,\,\tau=+1)$} & \quad\quad  & \multicolumn{3}{c}{$(\k,\,\tau=-1)$}\tabularnewline
			\cmidrule[0.4pt]{3-5}
			\cmidrule[0.4pt]{7-9}
			\multicolumn{1}{c}{locale} & \multicolumn{1}{c}{ \enskip op. \quad } & $\enskip$$-\Q$$\enskip$ & $\enskip$$ \phantom{+} \vec 0$$\enskip$ & $\enskip$$+\Q$$\enskip$ &  & $\enskip$$-\Q$$\enskip$ & $\enskip$$\phantom{+}\vec 0$$\enskip$ & $\enskip$$+\Q$$\enskip$\tabularnewline
			\midrule
			& $\hat{C}_{\mathrm{3A}}$ & $+1$ & $+1$ & $+1$ &  & $-1$ & $-1$ & $-1$\tabularnewline
			A & $\hat{C}_{\mathrm{3B}}$ & $\phantom{+}0$ & $-1$ & $+1$ &  & $-1$ & $+1$ & $\phantom{+}0$\tabularnewline
			& $\hat{C}_{\mathrm{3C}}$ & $-1$ & $\phantom{+}0$ & $+1$ &  & $-1$ & $\phantom{+}0$ & $+1$\tabularnewline
			\midrule[0.4pt]
			& $\hat{C}_{\mathrm{3A}}$ & $\phantom{+}0$ & $+1$ & $-1$ &  & $+1$ & $-1$ & $\phantom{+}0$\tabularnewline
			B & $\hat{C}_{\mathrm{3B}}$ & $-1$ & $-1$ & $-1$ &  & $+1$ & $+1$ & $+1$\tabularnewline
			& $\hat{C}_{\mathrm{3C}}$ & $+1$ & $\phantom{+}0$ & $-1$ &  & $+1$ & $\phantom{+}0$ & $-1$\tabularnewline
			\bottomrule \bottomrule
		\end{tabular}
		%\end{ruledtabular}
		\hspace*{\fill}
		\caption{Angular momentum quantum numbers $\ell^\mathrm{(s)}_{\k\tau}$ of momentum states relative to the three rotation centers. The condensate remains dark since all $\k=\vec{0}$ states have identical $\hat{C}_\mathrm{3C}$ eigenvalues of 0. \label{tab:angular_momentum}}
	\end{table}
	
	\subsection{Rotation symmetry}
	This robust darkness arises from rotation symmetry. Since both the condensate and light cones lie at high-symmetry $\k$ points, they can be simultaneous eigenstates of both linear and angular momentum, and thus the combined action of both translation and rotation symmetry will set optical selection rules.

	The three-fold $C_3$ rotational symmetry of the monolayers is lifted to the \moire{} superlattice as well. Like the monolayers, the bilayer  possesses multiple three-fold symmetries, here $\hat{C}_{3s}$ around the three locales $s=$ A, B, and C. 
	Excitons inherit the optical selection rules dictated by the local stacking configuration, determining the angular momenta  $\ell_{\mathrm{A}\tau} = +\tau$ and $\ell_{\mathrm{B}\tau} = -\tau$ for excitons localized in A and B locales, respectively \cite{Yu2017, WuLovornMacDonald2018}, and showing they are circularly polarized.
	
	Momentum states, however, will transform differently.
	The rotation operations around the three \moire{} locales are related by [cf. Fig.~\ref{fig:geometry}]
	\begin{equation} \label{eq:rotation_operations}
		\hat{C}_\mathrm{3B} = \hat{T}_{\vec{a}_1} \hat{C}_\mathrm{3A}, \qquad \hat{C}_\mathrm{3C} = \hat{T}_{\vec{a}_1} \hat{C}_\mathrm{3B}, 
	\end{equation}
	with $\hat{T}_{\vec{a}_1} = \exp (-i \hat{\vec{\mathcal{P}}} \cdot \vec{a}_1)$ an elementary lattice translation. 
	It follows that if a $\k$ state is an angular momentum eigenstate with angular momenta numbers $\ell^\mathrm{(A,B)}_{\k\tau}$ with respect to symmetries $\hat{C}_\mathrm{3A,3B}$, they must be related by
	\begin{align}
		e^{-\frac{2 \pi i}{3} \ell^\mathrm{(B)}_{\k\tau}} \xdop{C_3\k \tau} 
		& = \hat{C}_\mathrm{3B}^{\vphantom{\d}} \xdop{\k \tau} \hat{C}_\mathrm{3B}^\d  
		= \hat{T}_{\vec{a}_1}e^{-\frac{2 \pi i}{3} \ell^\mathrm{(A)}_{\k\tau}} \xdop{C_3\k \tau} \hat{T}_{\vec{a}_1}^\d
		\tcbr 
		= e^{-\frac{2 \pi i}{3} \ell^\mathrm{(A)}_{\k\tau} -i (\k-\tau\Q)\cdot\vec{a}_1} \xdop{C_3\k \tau}.
	\end{align}
	Conversely, $\k$ can be a rotational eigenstate only if $C_3 \k = \k$ up to reciprocal lattice vectors. This holds only at the high-symmetry $\k$ points, that is the zone center and corners. Writing $\k = \kappa \Q$, $\kappa = 0, \pm 1$, we thereby find
	\begin{equation}
		\ell^\mathrm{(B)}_{\k\tau} - \ell^\mathrm{(A)}_{\k\tau} =  \frac{\vec{a}_1 \cdot (\k - \tau \Q)}{2 \pi / 3} \!\!\!\mod 3 = -(\kappa - \tau) \!\!\!\mod 3.
	\end{equation}
	The same identity relates $\ell^\mathrm{(B)}_{\k\tau}$ and $\ell^\mathrm{(C)}_{\k\tau}$. 
	Using these relations, we list $\ell^\mathrm{(s)}_{\k\tau}$ of all high-symmetry momenta states formed by the lowest-energy orbitals in the A and B \moire{} sites in Table \ref{tab:angular_momentum}.

	We find that for the condensed $\k=\vec{0}$ modes, $\ell^{(s)}_{\vec{0}\tau}$ depend on the rotation center. This indicates rotation symmetry also prevents their recombination, as they are symmetry-distinguishable from photons, which have uniform angular momentum with respect to all three centers. 
	
	Yet like translation symmetry, the BEC invalidates rotation centers $s$ for which  $\ell^{(s)}_{\vec{0}\tau}\neq \ell^{(s)}_{\vec{0}\bar{\tau}}$, potentially enabling emission. 
	We find that $\hat{C}_\mathrm{3A,3B}$ are broken, while $\hat{C}_\mathrm{3C}$ survives with  $\ell^{(\mathrm{C})}_{\vec{0}\tau} =  \ell^{(\mathrm{C})}_{\vec{0}\bar{\tau}} = 0$. 
	While  only one $\ell$ now characterizes $\k=\vec{0}$ modes, this vestigial symmetry prevents emission: To conserve energy, high-symmetry excitons must emit a photon perpendicular to the bilayer, i.e. with $\ell=\pm1$, and therefore $\ell_{\vec{0}\tau}^{(\mathrm{C})}=0$  dictates that the condensate remains dark. This selection rule is protected by the time-reversal duality between valleys and the high symmetry of the momentum mismatch $\Q$.
	
	A similar situation where momentum-folded states remain optically dark was observed in  Smolensky et al., see Ref.~\cite{Smolenski2021}.
	In that experiment, translation symmetry is broken by the charge-ordered Wigner crystallization of doped electrons, with wavenumber $\k_w$. 
	Since $|\k_w| < |\Q| $, six degenerate, low-symmetry exciton states at $\p=C_6^n \k_w$ are folded on top of each other and the light cone. 
	The six states then hybridize and redistribute into states of well-defined angular momentum. 
	Of those, only two have $\ell = \pm 1$ and are allowed to couple to the light cone and become bright, while four states have different $\ell$ and thus remain dark. 
	In our case, the commensurability of the exciton density wave leads to only two independent states $\p=\pm \K$ being folded, which are themselves $\ell=0$ eigenvalues, and therefore we do not observe the same richness here.
	
	In total, we find that in these twisted bilayers, two-valley condensates lie on the verge of optical activation, which is driven by valley interconversion, and stopped by a vestigial rotation symmetry. Therefore, bright BECs could be achieved by applying rotation symmetry breaking fields and strain, and further enhanced by out-of-plane pressure which increases valley-depolarizing e--h exchange interactions. We leave this, and other interesting intervalley phenomena, to future work.
	
	Finally, we remark that recombination of $\k\neq\vec{0}$ excitons via momentum folding remains possible due to their low rotation symmetry. However, this presents a secondary incoherent emission channel that is dominated by the leaky emission we consider in the main text. We demonstrate this below.

	\section*{Intervalley excitons}
	In our Letter we study the properties of excitons formed of electrons and holes in the same K valley of the constituent monolayers. These are known as intravalley, or K--K, excitons. Thus, the bandgap is rendered indirect by the interlayer twist. This holds, for example, in {MoS$_2$/WSe$_2$} \cite{Karni2019, Tan2021, Karni2021}, and our treatment applies directly.
	
	For some combinations of materials and stacking configurations, hybridization between the two layers can lead to an indirect bandgap even at zero twist, subtended by different $\k$ points such as $\Gamma$--K, $\Gamma$--$\Lambda$, and K--$\Lambda$ \cite{Hagel2021}. Thus, momentum-dark intervalley excitons  composed of electrons and holes in different valleys could be lower in energy, and would be those to condense. Their intervalley momentum mismatch is too large to be compensated by center-of-mass motion; thus, while strong dipole--dipole interactions may play a part in facilitating recombination, our theory does not apply to them.
	
	However, our theory may  nevertheless still apply depending on the relevant timescales: since excitons are generated optically, intravalley excitons are formed first. The extremely long valley depolarization times observed in experiment, up to hundreds of ns \cite{Seyler2019,Jiang2018}, would suggest the conversion between intra- and intervalley excitons is slow, and momentum-direct excitons will be persistent. For example, intravalley excitons still lead to pronounced photoluminescence  in MoSe$_2$/WSe$_2$ heterobilayers \cite{Nayak2017, Nagler2019, Tran2019}. Therefore, it is feasible that over experimentally-relevant timescales comparable to our $\tau_\mathrm{loc}$, we can consider a population of photogenerated intravalley excitons, and our theory remains applicable.
	
	\section*{The Rotating Frame} \label{app:RWA}
	
	To study emission from condensed states, we must consider populated, i.e. excited states of Hamiltonian \eqref{eq:single_valley_hamiltonian}. 
	Nevertheless, it is convenient to move to a frame in which the condensate is the ground state, so that various observables are given by ground state expectation values. 
	Such a transition is implemented by the unitary operator $e^{i \mu \hat{\mathcal{N}} t}$ with $\hat{\mathcal{N}}$ the particle counting operator. 
	However, this would lead to a time-dependent light--matter interaction $\hat{V}_\mathrm{LMI}(t)$ in the new frame. This is avoided by having $\hat{\mathcal{N}}$ count photons as well. 
	The Hamiltonian is then transformed into
	\begin{equation}
		\hat{\Xi} = \hat{H}- \mu \hat{\mathcal{N}},\qquad 
		\hat{\mathcal{N}} = \sum_\k \xdop{\k}\xdop{\k}+\sum_{\p,\sigma} \adop{\p \sigma}\aop{\p \sigma},
	\end{equation}
	and  $\hat{\Xi}$ is time-independent since $[\hat{H},\, \hat{\mathcal{N}}]=0$ thanks to the rotating-wave approximation in Eq.~\eqref{eq:V_LMI}. 
	$\hat{\Xi}$ then appears as a standard excitonic grand-canonical potential with chemical potential $\mu$, perturbatively coupled to photons with energies $(\hbar\omega_\p - \mu)$. As this frame allows us to focus on the low-energy physics of the condensate, this shift sets the emission spectrum  around the bilayer optical bandgap $\mu \approx E_0 $.

	\section*{Hard-Core Bosons and Spin Waves} \label{app:hard-core_bosons}
	
	In this Section we briefly outline the analysis of the hard-core boson and corresponding spin model studied in the main text.
	
	\subsection{Particle--hole symmetry}
	The hard-core constraint gives rise to the notion of boson vacancies. A particle--hole-like symmetry $\hat{\mathcal{P}}\sop{\R}^\pm \hat{\mathcal{P}} = \sop{\R}^\mp$ connects the ground state of Eq.~\eqref{eq:XXZ-Hamiltonian} with filling $\nu$ to that with filling  $2S-\nu$ (here $S=\tfrac{1}{2}$ throughout). Therefore, one can show
	\begin{equation} \label{eq:particle-hole-relation}
		n_\k (\nu) - n_\k (2S-\nu) 
		%= \Braket{\left[\sop{\k}^+,\sop{\k}^-,\right]} 
		= 2\braket{\sop{}^z}=2 \nu - 2S.
	\end{equation}
	This relation places a strong restriction on the resulting emission rate. For a finite system size $N$,  $n_\k (0)=n_{\k\neq0} (1/N) = 0$, leading to the leading-order dependences at low and high filling
	\begin{equation} \label{eq:particle-hole-asymptotes}
		n_{\k\neq0} (\nu) = \begin{cases}
			0 + \mathcal{O}(\nu^2),  & \nu \to 0 \\ 
			2\nu - 2S + \mathcal{O}((2S-\nu)^2), & \nu \to 2S.
		\end{cases}
	\end{equation}
	These two asymptotes can be satisfied simultaneously only by a nonlinear dependence, thus establishing the role of interactions.  
	
	\subsection{Mean-field theory}
	The first refinement to Eq.~\eqref{eq:particle-hole-asymptotes} is found in classical mean-field theory. We write the mean-field wavefunction 
	\begin{equation} \label{eq:mean_field_wavefunction}
		\Ket{\Psi_{\mathrm{MF}}}=
		\bigotimes_{\R} \left(\cos \tfrac{\vartheta}{2} \ket{\Uparrow} + \sin \tfrac{\vartheta}{2} \ket{\Downarrow}\right)_\R
	\end{equation}
	which corresponds a uniform polarization in the XZ plane at polar angle $\vartheta$, and is the mean-field ground state for $\mu = E_0 + z \cos \vartheta$, where $z=6$ is the lattice coordination number. At mean-field level the filling is $\nu_\mathrm{MF} = \braket{S^z} + S = S \left(\cos \vartheta + 1\right)$. A straightforward computation yields
	\begin{equation}
		n_{\k\neq\vec{0}} (\nu_\mathrm{MF}) = \braket{\sop{\k}^+ \sop{\k}^-} = \cos^4 \tfrac{\vartheta}{2} = \nu_\mathrm{MF}^2,
	\end{equation}
	which is consistent with Eq.~\eqref{eq:particle-hole-asymptotes}. The factorized wavefunction \eqref{eq:mean_field_wavefunction} is totally uncorrelated. However, due to their strong repulsion, particles should be \emph{anti}correlated, leading to fewer scatterings and thus weaker emission. Therefore, this result is an upper bound on the emission rate. In addition, its independence of $\k$ motivates us to seek the next order correction.
	
	\subsection{Spin-wave theory} 
	Our treatment closely follows that of Bernardet et al \cite{Bernardet2002} and references therein with suitable adaptations to the present lattice geometry.  We first rotate the spins in the XZ plane such that the mean-field state points in the negative $z$ direction, and then perform a Holstein-Primakoff (HP) $1/S$ expansion \cite{HolsteinPrimakoff1940}. We substitute for the spin fields
	\begin{align}
		\hat{S}_\R^x &= \phantom{-}\frac{\cos \vartheta }{2}\sqrt{2S \!-\! \hdop{\R}\hop{\R}}\hop{\R} + \frac{\sin \vartheta}{2} (S \!-\! \hdop{\R}\hop{\R}) + \hc, \nonumber \\
		\hat{S}_\R^z &= -\frac{\sin \vartheta }{2}\sqrt{2S \!-\! \hdop{\R}\hop{\R}}\hop{\R} + \frac{\cos \vartheta}{2} (S \!-\! \hdop{\R}\hop{\R}) + \hc, \nonumber \\
		\hat{S}_\R^y & = \frac{1}{2i}\sqrt{2S \!-\! \hdop{\R}\hop{\R}}\hop{\R} + \hc
	\end{align} where $\hop{\R}$ are the HP bosonic annihilation operators.
	
	Linearizing results in a quadratic Hamiltonian that is diagonalized by the mode expansion 
	$\hat{\Xi}=\sum_\k \Omega_{\k} \bdop{\k} \bop{\k}$, with dispersion $\Omega_\k^2 = \epsilon_\k (\epsilon_\k \cos^2\vartheta + z t \sin^2\vartheta)$ where $\epsilon_\k$ is the dispersion in the main text. In terms of these collective modes, the real exciton annihilation operator is
	\begin{align} \label{eq:hard-core_bosons_emission_operators}
		\xop{\k\neq \vec{0}} & = \left(
		\cos^2 \tfrac{\vartheta}{2} \cosh \theta_\k  +
		\sin^2 \tfrac{\vartheta}{2} \sinh \theta_\k
		\right) \bdop{-\k} 
		\tcbr
		- 
		\left(
		\sin^2 \tfrac{\vartheta}{2} \cosh \theta_\k  +
		\cos^2 \tfrac{\vartheta}{2} \sinh \theta_\k
		\right) \bop{\k} 
		+\dots
	\end{align} 
	The BdG mixing angles are again given by the rhs of Eq.~\eqref{eq:bogoliubov_modes}, writ explicitly
	\begin{equation} \label{eq:hardcore-bosons-mixing-angles}
		\sinh^2 \theta_\k   = \frac{1}{2}\left(\frac{\left(1+\cos^{2}\vartheta\right)\epsilon_{\k}+  zt\sin^{2}\vartheta}{2\Omega_{\k}}-1\right).
	\end{equation}
	
	The ellipses in Eq.~\eqref{eq:hard-core_bosons_emission_operators} represent sub-leading terms that are higher powers of $\bop{},\,\bdop{}$. The HP picture naturally shows that higher order processes that leave behind more than one collective mode will appear in the theory. Yet these contributions to $n_\Q$ cannot be evaluated consistently without computing higher-order corrections to the ground state as well, so are dropped.
	
	Under the same approximation of a nearly-uniform occupancy inside the light cone, we find
	\begin{align}
		\Gamma(\nu) & =  N \tau_\mathrm{loc}^{-1} \times 
		\left(\cos^2 \tfrac{\vartheta}{2} \cosh \theta_\Q  -
		\sin^2 \tfrac{\vartheta}{2} \sinh \left| \theta_\Q\right|
		\right)^2  \tcbr
		=  N \tau_\mathrm{loc}^{-1} \times \left[\frac{\cos \vartheta}{2}+\frac{2+\cos 2 \vartheta}{\sqrt{30 +6\cos 2 \vartheta}} \right].
	\end{align}
	Note that $\theta_\k < 0$ if $\epsilon_\k > zt$, and here we have explicitly used that it is negative at $\Q$. This shows suppression of emission compared to the mean-field result. 
	$\Gamma$ may be expressed explicitly by radicals, and in the dilute limit
	\begin{equation}
		\Gamma(\nu \ll 1) \approx  N \tau_\mathrm{loc}^{-1} \left(\frac{z t}{\epsilon_\Q}\right)^2 \cos^4 \tfrac{\vartheta}{2} 
		= \frac{4}{9}  N \tau_\mathrm{loc}^{-1} \nu_\mathrm{MF}^2.
	\end{equation}
	
	Truncation of the HP expansion generically breaks Hamiltonian symmetries \cite{Vogl2021}, and here the particle--hole relation \eqref{eq:particle-hole-relation} is violated. However, trigonometric identities show that $\Gamma(\nu)$ upholds \eqref{eq:particle-hole-relation} at the level of the mean-field density $\nu_\mathrm{MF}$. Except for very small fillings (see below), the numerical correction between $\nu$ and $\nu_\mathrm{MF}$ is merely quantitative, and we neglect it when evaluating $\Gamma(\nu)$ in the main text. This recovers the particle--hole symmetry of $\Gamma(\nu)$, e.g. in Fig.~\ref{fig:hard-core_boson}.

	\subsection{Nonzero temperature}
	Thermal excitations will also lead to a population of zone-corner excitons, and are the primary channel considered for exciton loss, e.g. Ref.~\cite{Choi2021}. In terms of spin waves, this corresponds to processes wherein thermal collective modes are absorbed, rather than emitted, to sink excess momentum. 
	
	Thus, at $T>0$ two emission lines emerge around $E_0$,
	\begin{equation}
		\hbar \omega = \mu \mp  \Omega_{\Q} = E_0 + z t (\cos \vartheta\mp\sqrt{r^2 \cos^2 \vartheta + r \sin^2 \vartheta}),
	\end{equation}
	where $r = \epsilon_\Q / zt$ (=3/2) and the signs correspond to processes that emit and absorb a spin wave, respectively. 
	Interactions lead to a blueshift with increasing density,
	\begin{equation}
		\hbar \omega_\mathrm{emit} \approx E_0 -\frac{5}{2}zt + 3zt \nu, \; 
		\hbar \omega_\mathrm{absorb}  \approx E_0 + \frac{1}{2} zt + zt \nu.
	\end{equation}
	The two emission peaks are split by $3zt\sim20t$.  For a representative $t \sim 0.2~\mathrm{meV}$  \cite{Yu2017}, this separation is $\sim 4 \mathrm{meV}$, and might be resolved spectroscopically.

	The matrix element for the two  processes are given by the parentheses in Eq.~\eqref{eq:hard-core_bosons_emission_operators}. Unlike the usual Stokes and anti-Stokes lines in phonon-assisted emission \cite{MoskalenkoSnoke2000}, they are not equal. Rather, at high temperature the ratio of intensities reproduces the leaky condensate scaling
	\begin{align}
		\frac{\Gamma_-}{\Gamma_+} 
		\approx \left[\frac{ 
			\cos^2 \tfrac{\vartheta}{2} \cosh \theta_\k  +
			\sin^2 \tfrac{\vartheta}{2} \sinh \theta_\k}
		{\sin^2 \tfrac{\vartheta}{2} \cosh \theta_\k  +
			\cos^2 \tfrac{\vartheta}{2} \sinh \theta_\k }\right]^2 \!\!\!
		\approx \frac{4}{9} \nu_\mathrm{MF}^2
	\end{align}
	and the anti-Stokes-like line dominates at high $T$. In contrast, Bogoliubov theory \eqref{eq:bogoliubov_modes} gives $\Gamma_-/\Gamma_+ \approx \tanh^2  \theta_\Q \approx \left(\nu_c U / 2 \epsilon_\Q \right)^2$.
	Therefore, the ratio of intensities could be also used to estimate the strength of exciton interactions.
	
	The combined contribution of both processes to the zone-corner exciton  population is
	\begin{align}
		n_{\Q} (T)  &=
		n_\Q (0) + \frac{3 - 2 \sin^2 \vartheta }{\sqrt{9-3 \sin^2 \vartheta }} \times \frac{1}{e^{\Omega_{\Q}/k_{B}T}-1}\,.
	\end{align}
	The first and second terms correspond to the contributions of interactions at $T=0$ and thermal excitations, respectively. 
	It is then natural to ask which of the mechanisms is dominant. 
	The inset of Fig.~\ref{fig:hard-core_boson} depicts the crossover temperature $T^\star$ at which the two make equal contributions.

	The natural temperature scale to compare against is the Berezinskii--Kosterlitz--Thouless (BKT) superfluid phase transition temperature, given by \cite{KosterlitzThouless1973} 
	\begin{equation} \label{eq:BKT_critical_temperature}
		kT_\mathrm{BKT} = \frac{\hbar^2}{m}\frac{\pi}{2} n_s  \apprle \sqrt{3} \pi \nu t.
	\end{equation}
	Here $n_s = \nu_s / \mathcal{A}_M \le \nu / \mathcal{A}_M$, with $\mathcal{A}_M $ the \moire{} supercell area, is the number density of the superfluid component at $T=0$, and in the effective mass approximation for $\epsilon_\k$, $\hbar^2/m = 2\sqrt{3} \mathcal{A}_M t $. Similarly, we estimate the degeneracy temperature by $kT_d = 2 \pi \hbar^2 n / m \approx 4\sqrt{3}\pi \nu t$.
	Fig.~\ref{fig:hard-core_boson} shows that at all densities $T^\star > T_\mathrm{BKT}$, demonstrating that interactions dominate the emission in the superfluid phase and much of the Bose-degenerate regime. 
	
	\subsection{Numerical results}
	We compare our HP expansion against exact diagonalization by implementing the Hamiltonian \eqref{eq:XXZ-Hamiltonian} numerically.  We set $\mu = E_0$ for convenience, and compute numerically the ground state of $\hat{\Xi}$  in each of the fixed density sectors $\nu = 2/N,\,3/N \dots 1$ separately. 
	We then evaluate $\braket{\hat{S}_\Q^+ \hat{S}_\Q^-}$ in this state to obtain $n_\Q(\nu)$ of each sector. 
	Commensurability of $\Q$ with periodic boundary conditions limits the triangular crystal dimensions to be $N=3n\times3m$ with $n,m$ integers.
	Fig.~\ref{fig:hard-core_boson} shows the results for a $6\times3$ lattice.
	
	\begin{figure*}[tb]
		\includegraphics[width=\apsfigurewidth]{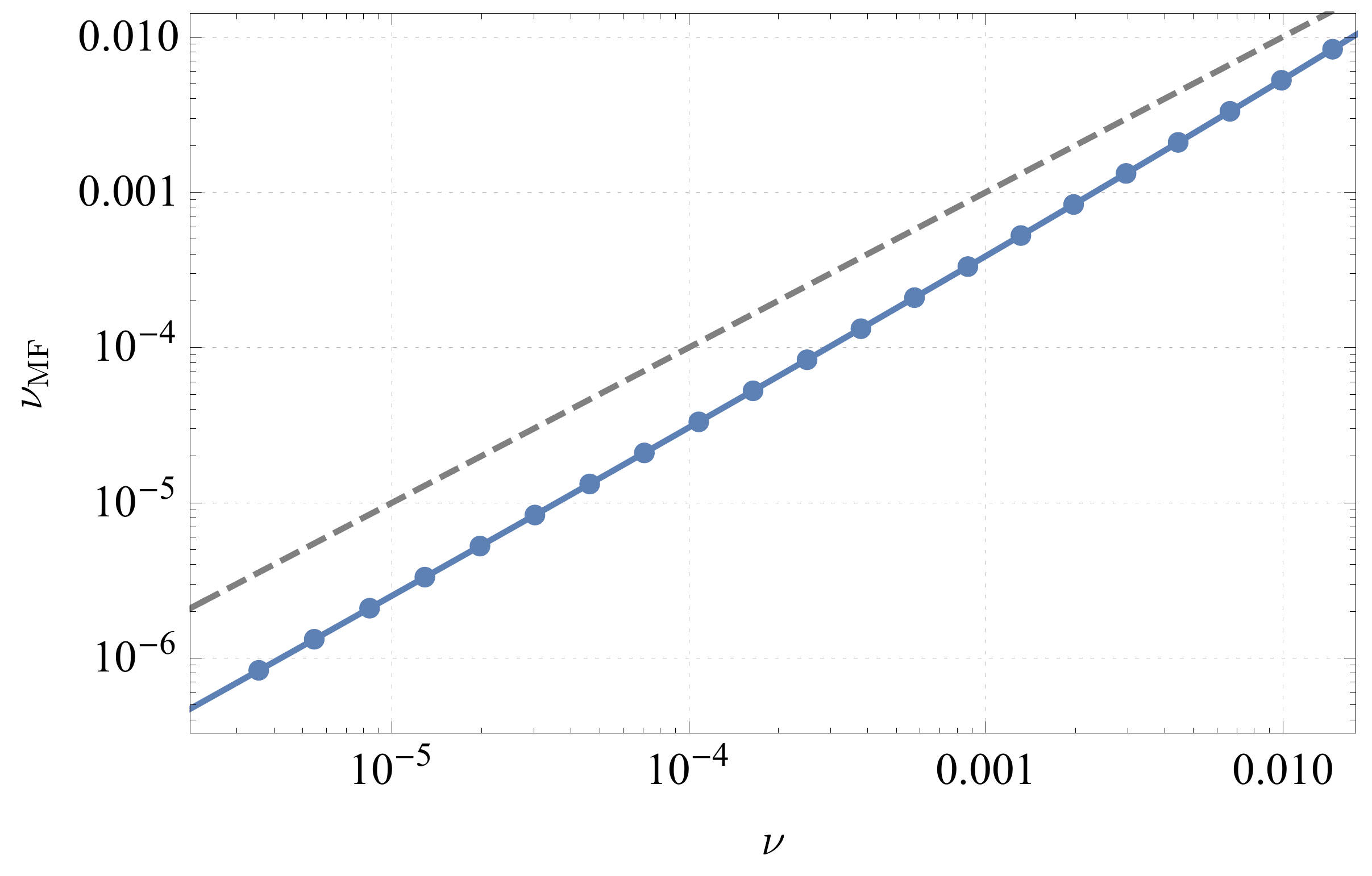}
		\includegraphics[width=\apsfigurewidth]{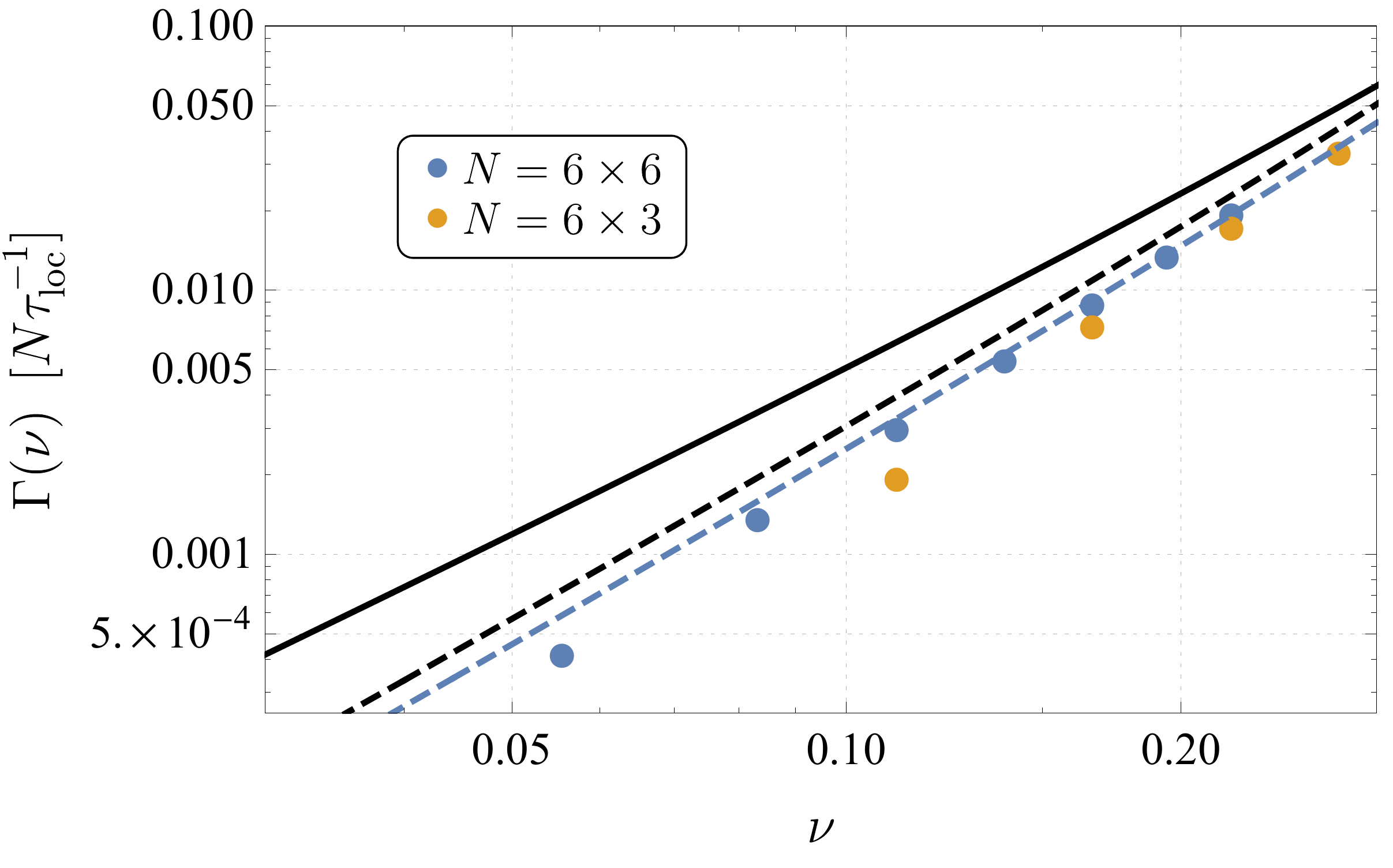}
		\caption{
			Left: Mapping between $\nu$ and $\nu_\mathrm{MF}$. The dots are calculated numerically, and the solid line is a fit to $\nu_\mathrm{MF} \propto -\nu/\log(\nu/\nu_0)$, giving excellent agreement with $\nu_0=5.0$. The dashed line marks $\nu_\mathrm{MF} = \nu$ for comparison. 
			Right: Asymptotics of emission rate at low filling. The emission rate $\Gamma(\nu)$ is plotted without (solid line, same as in Fig.~\ref{fig:hard-core_boson}) and with (dashed black line) spin-wave density corrections between $\nu_\mathrm{MF}$ and $\nu$. 
			Focusing on smaller filling allows us to exactly diagonalize a larger $6\times6$ system, which we fit with the functional form \eqref{eq:asymptotic_fit} (dashed blue line). 
			Note the spin-wave corrections give the correct asymptotic form for $\nu\to0$. The remaining discrepancy could be due to finite-size effects.
		}
		\label{fig:asymptotics}
	\end{figure*}
	
	\subsection{Asymptotic correlations}
	We would like to estimate the emission rate for very low filling. In this regime, correlation lengths become large, yet $n$-particle correlations are suppressed by factors of $\nu^n$. Therefore,  2-particle correlations are dominant \cite{MazuzHarpaz2019}.  We thus consider a fixed number of $N_s = 2$ hard-core bosons, while taking the system size $N\to\infty$ to scan  $n_\Q(\nu)$ for $\nu=\frac{2}{N}\to0$. The order of thermodynamic limits is important, and we expect different results for $n_\Q(\nu)$ if $N_s$ or $\nu$ are held fixed while taking $N \to \infty$, and we therefore use this to extract scaling laws without precise numerical factors.

	The two-body ground state with energy $E_\mathrm{GS}$ is generally written
	\begin{equation}
		\Ket{\mathrm{GS}} = \sqrt{\frac{2}{N}}\sum_{\R, \vec{\Delta}} \psi_\vec{\Delta} \sop{\R + \vec{\Delta}}^+ \sop{\R }^+ \Ket{\mathrm{vac}}
	\end{equation}
	where $\psi_\vec{\Delta}$ is a real and symmetric normalized two-body pairing wavefunction, with $\psi_\vec{0} = 0$ due to the hard-core constraint. The zone-corner occupation is then
	\begin{equation}
		n_\Q = 2\times \left|\tilde{\psi}_\Q\right|^2
	\end{equation}
	where $\tilde{\psi}_\k = \sum_\vec{\Delta} e^{-i\k\cdot\vec{\Delta}} \psi_\vec{\Delta} / \sqrt{N}$.

	Substituting this state into the Hamiltonian  \eqref{eq:XXZ-Hamiltonian} gives a discretized Schrodinger equation for $\psi$, solved by 
	\begin{equation}
		\left|\tilde{\psi}_\k\right|^2 = \frac{1}{(\epsilon_\k - \varepsilon)^2} 
		\Bigg/ {\sum_{\k'}  \frac{1}{(\epsilon_{\k'}-\varepsilon)^2} }.
	\end{equation}
	Here $\varepsilon = E_\mathrm{GS}/2 - \epsilon_{\vec{0}}$ is the mean energy of each particle relative to the non-interacting single-particle band minimum, determined implicitly by the hard-core condition
	\begin{equation}
		\psi_\vec{0} \propto \sum_\k  \frac{1}{\epsilon_\k-\varepsilon} = 0\,. 
	\end{equation}
	
	So far this solution is exact. We proceed to evaluate $n_\Q$ asymptotically by
	\begin{align}
		n_\Q (\nu) & 
		\sim \frac{\varepsilon^2}{\epsilon^2_\Q}
		\sim 
		{\left[\sum_{\k \neq \vec{0}}  \frac{\epsilon_\Q}{\epsilon_\k} \right]^{-2} } 
		\sim \left[ N \int \frac{k dk}{k^2} \right]^{-2} \tcbr
		\sim \left[ - N \log {N/N_0} \right]^{-2} \sim \frac{A \nu^2}{(\log(\nu / \nu_0))^2}
	\end{align}
	where $A$ is a proportionality constant, and $N_0$ and $\nu_0$ are some positive constants associated with an infrared cutoff of the integral's logarithmic divergence at the origin, corresponding to omitting $\k=\vec{0}$ from the initial sum. 
	
	While this relation was derived for a two-particle problem, it provides a remarkably good description of $n_\Q(\nu)$ for more particles, as seen in Fig.~\ref{fig:hard-core_boson}. Demanding a smooth transition at the $\nu=\frac{1}{2}$ particle--hole symmetry point with $dn_\Q/d\nu = 1$ eliminates $A$, and we obtain the form 
	\begin{equation} \label{eq:asymptotic_fit}
		n_\Q(\nu < \tfrac{1}{2}) = \frac{[\ln \, 2 \nu_0]^3}{1 +\ln \, 2 \nu_0 }
		\frac{\nu^2 }{[\ln\nu/\nu_0]^2}
	\end{equation}
	with appropriate symmetrization for $\nu>\frac{1}{2}$, and a single fitting parameter $\nu_0$ (to which we ascribe no physical meaning). Fitting numerical results with Eq.~\eqref{eq:asymptotic_fit}, we obtain $\nu_0 \sim 5.5$. 
	
	Our spin-wave result Eq.~\eqref{eq:hard-core_bosons_emission_rate} is consistent with Eq.~\eqref{eq:asymptotic_fit} only if $\nu_\mathrm{MF}\sim -\nu/\log(\nu/\nu_0)$ with a similar $\nu_0$. Following Bernardet et al \cite{Bernardet2002}, we numerically evaluate the density from the defining thermodynamic relation $\nu = -\frac{1}{N}\frac{\partial}{\partial \mu} \braket{\hat{\Xi}}$. $\nu$ then differs from $\nu_\mathrm{MF}$ by the contribution of the spin-wave zero-point motion to the ground state expectation value $\braket{\hat{\Xi}}$.  We plot $\nu_\mathrm{MF}$ versus $\nu$ in Fig.~\ref{fig:asymptotics}a, and fit the expected logarithmic correction to excellent agreement, with $\nu_0 \sim 5.0$. The close agreement between the two fitted values shows that the spin-wave theory correctly captures the correlations at small filling. We then plot the emission rate in terms of the full density in Fig.~\ref{fig:asymptotics}b, showing good agreement between the spin-wave calculation and exact diagonalization data.
	
	\section*{Emission at small $\textbf{k}$}

	Above we show that the coupling between $\k=\vec{0}$ condensate modes and the density wave is forbidden by $\hat{C}_3$ symmetry. Nonzero $\vec{k}$ modes have lower symmetry, and therefore may couple to the density wave to recombine. This is another interaction-driven incoherent emission channel that is nonlinear in exciton density, and therefore acts as an additional leaky emission channel, further reinforcing the dominance of interactions at low temperatures. Here we show its contribution is subleading compared to the primary channel we explore in the main text.
	
	While this effect requires a  two-valley model, results of our hard-core theory are still useful for estimating its magnitude. So far, leaky emission was facilitated by spontaneous creation of zone-corner Bogoliubov modes, which mix the creation and annihilation of real excitons. The momentum folding induced by the density wave allows zone-center Bogoliubov mode creation $\bdop{-\k}$ to mix an annihilation of a bright zone corner exciton $\xop{\k+\Q}$ as well. Therefore, emission via this channel is given by $\tilde{\Gamma} = \sum_\k \sinh^2 \theta_\k \braket{\xop{\k+\Q}\bop{-\k}} \Gamma_{\k+\Q}$, where the first two factors are the squared amplitude for spontaneous production of the BdG mode, and its bright fraction, respectively. (The latter is unity for zone-corner emission.) The summation runs over the folded light cone $|\k| \le E_0 / \hbar c$.
	
	We now estimate all factors for small $\k$. From Eq.~\eqref{eq:hardcore-bosons-mixing-angles},  $\sinh^2 \theta\approx \Omega_\k / 2 \epsilon_\k \sim \sqrt{\nu / (k a)^2}$. As we show in our symmetry analysis, a genuine momentum folding density wave necessitates intervalley conversion. Thus, $\braket{\xop{\k+\Q}\bop{-\k}}\sim |\nu U J^\mathrm{IV}_{\k+\Q} / \epsilon_{\Q}^2|^2$ which for strong interactions saturates to $|\nu  J^\mathrm{IV}_{\k+\Q} / \epsilon_{\Q}|^2$. Finally, the $\hat{C}_3$ forbidden transition at $\k=0$ [compare $\ket{\k=\vec{0}, \tau=+1}$ and $\ket{\k = \vec{-\Q},\tau=-1}$ in Table~\ref{tab:angular_momentum}] manifests in the suppressed  $|J^\mathrm{IV}_{\k+\Q}| \sim J_0 \times a k$ around the zone center, where $J_0$ is the typical amplitude for intervalley conversion.
	
	Collecting contributions, in total we find
	\begin{equation}
		\tilde{\Gamma} \!
		\sim \! N \!\! \int_0^{\frac{1}{\lambda}} \!\! \frac{k dk}{a^{-2}} \frac{J_0^2}{\epsilon_\vec{Q}^2} a k \nu^{\tfrac{5}{2}}
		\Gamma_{\k+\Q} \!
		\sim N \tau_\mathrm{loc}^{-1}  \frac{a}{\lambda} \frac{J_0^2}{\epsilon_\vec{Q}^2}  \nu^{5/2}
		%\frac{\Gamma_\vec{Q}}{\sqrt{1-k^2 \lambda^2}} 
		\sim \frac{a}{\lambda} \frac{J_0^2}{\epsilon_\vec{Q}^2} \! \sqrt{\nu} \times  \Gamma.
	\end{equation}
	Here $\Gamma$ is the leaky emission rate computed in the main text. 
	$\lambda \sim \hbar c / E_0 \sim 1~\mathrm{\mu m}$ is the optical wavelength.
	Furthermore, that intervalley conversion is driven by e--h exchange interactions that also determine recombination allows us to estimate $J_0 \approx \mathcal{D}^2/(4\pi\epsilon a^3)\sim0.01$ meV,  where $\mathcal{D}$ is the dipole transition matrix element, about 0.5 $e\AA$ \cite{WuLovornMacDonald2018}. We thus have $a/\lambda \approx 10^{-2}$ and $J_0/\epsilon_\Q \sim 10^{-2}$--$10^{-1}$, showing this emission channel is substantially weaker than that of the main text.

	We remark that vertically polarized excitons (e.g. Ref.~\cite{Wang2017}) are not $\hat{C}_3$-forbidden from participating in the momentum folding, and in that case $|J^\mathrm{IV}_{\k+\Q}| \sim J_0 $. However, the inner product between this polarization and the outgoing photon polarization reintroduces a $k^2$ scaling, leading to the same result as above. Moreover, while such exciton polarization is observed in monolayers, it might not readily realize in bilayers \cite{Yu2018, Sigl2021}.

	%TC:endignore 
\end{document}